# Attosecond VUV Coherent Control of Molecular Dynamics

P. Ranitovic[1,4], C. W. Hogle[1], P. Rivière[3], A Palacios[3], X. M. Tong[2], N. Toshima[2], A. González-Castrillo[3], L. Martin[1], F. Martín[3,5], M. M. Murnane[1], and H. C. Kapteyn[1]

[1]*JILA and Department of Physics, University of Colorado and NIST, Boulder, CO 80309, USA*

[2]*Division of Materials Science, Faculty of Pure and Applied Science, University of Tsukuba, Ibaraki 305-8573, Japan*

[3]*Departamento de Química, Universidad Autónoma de Madrid, 28049 Madrid, Spain*

[4]*Lawrence Berkeley National Lab, 1 Cyclotron Road, Berkeley CA, 94720*

[5]*Instituto Madrileño de Estudios Avanzados en Nanociencia, IMDEA-Nano, Cantoblanco, 28049 Madrid, Spain*

## Abstract

High harmonic light sources make it possible to access attosecond time-scales, thus opening up the prospect of manipulating electronic wave packets for steering molecular dynamics. However, two decades after the birth of attosecond physics, the concept of attosecond chemistry has not yet been realized. This is because excitation and manipulation of molecular orbitals requires precisely controlled attosecond waveforms in the deep ultraviolet, which have not yet been synthesized. Here, we present a novel approach using attosecond vacuum ultraviolet pulse-trains to coherently excite and control the outcome of a simple chemical reaction in a deuterium molecule in a non-Born Oppenheimer regime. By controlling the interfering pathways of electron wave packets in the excited neutral and singly-ionized molecule, we unambiguously show that we can switch the excited electronic state on attosecond timescales, coherently guide the nuclear wave packets to dictate the way a neutral molecule vibrates, and steer and manipulate the ionization and dissociation channels. Furthermore, through advanced theory, we succeed in rigorously modeling multi-scale electron and nuclear quantum control in a molecule for the first time. The observed richness and complexity of the dynamics, *even in this very simplest of molecules*, is both remarkable and daunting, and presents intriguing new possibilities for bridging the gap between attosecond physics and attochemistry.

## Introduction

The coherent manipulation of quantum systems on their natural timescales, as a means to control the evolution of a system, is an important goal for a broad range of science and technology including chemical dynamics and quantum information science. In molecules, these timescales span from attoseconds timescales characteristic of electronic dynamics, to femtosecond timescales characteristic of vibrations and dissociation, to picosecond timescales characteristic of rotations in molecules. With the advent of femtosecond lasers, observing the transition state in a chemical reaction (*1*) and controlling the reaction itself, became feasible. Precisely-timed femtosecond pulse sequences can be used to selectively excite vibrations in a molecule, allow it to evolve, and finally excite or de-excite it into an electronic state not directly accessible from the ground state (*2*). Alternatively, interferences between different quantum pathways that end up in the same final state can be used to control the outcome of a chemical reaction (*3*-*9*).

In recent years, coherent high harmonic sources with bandwidths sufficient to generate either attosecond pulse trains or a single isolated attosecond pulses have been developed, that are also perfectly synchronized to the driving femtosecond laser (*10*-*12*). This new capability provides intriguing possibilities for coherently and simultaneously controlling *both* the electronic and nuclear dynamics in a molecule in regimes where the Born-Oppenheimer approximation is no longer valid, to select specific reaction pathways or products. Here, we realize this possibility, in a coordinated experimental-theoretical study of dynamics in the simplest neutral molecule: deuterated hydrogen ($D_2$).

The hydrogen molecule, as the simplest possible neutral molecule that can be fully described theoretically, has been the prototype molecule for understanding fundamental processes that lie at the heart of quantum mechanics (*13*-*16*). However, in such a small molecule, the coupled electron-nuclear dynamics are in the attosecond to few-femtosecond regime, while the electronically excited states lie in the vacuum ultraviolet (VUV) region of the spectrum. Because of the challenge of generating attosecond VUV waveforms using traditional laser frequency doubling or tripling in nonlinear crystals, it has not been possible to date to explore the dynamics of an electronically excited hydrogen molecule. Exploiting attosecond physics, on the other hand, only a handful of time-resolved experiments have been performed on $H_2$ ($D_2$), focusing mainly on controlling dissociation through electron localization in electronically excited $D_2^+$ ions (*17-23*). Recently it was realized that infrared (IR) femtosecond laser pulses, in combination with a phased locked comb of attosecond VUV harmonics, can be used to control the excitation and ionization yields in He on attosecond timescales, by interfering electron wave packets (*24*, *25*). These experiments extended the Brumer-Shapiro (3-9) two-pathway interference coherent control concept to the attosecond temporal and VUV frequency domain. More recently, it was shown that by manipulating the individual amplitude of VUV harmonics, it is possible to induce full electromagnetic transparency in He, by destructively interfering two electronic wave packets of the same amplitude and opposite phases (*26*). Other recent work used shaped intense femtosecond laser pulses to manipulate populations by controlling the oscillating charge distribution in a potassium dimer (*27*).

In this paper, we demonstrate for the first time that we can coherently and simultaneously manipulate multi-state electronic and multi-potential-well nuclear wave-packet dynamics, to control the excitation, nuclear wave packet tunneling, as well as the dissociation and ionization channels of the only electronically excited neutral molecule where full modeling of the coupled quantum dynamics is possible – $H_2$ (for practical reasons we used deuterated hydrogen in this experiment). By combining attosecond pulse trains of VUV with two near-IR fields, together with strong-field control that exploits a combination of the two-pathway interference Brumer-Shapiro (3-9) and pump-dump Tannor-Rice (2) approaches, we demonstrate that we can selectively steer the ionization, vibration and dissociation of $D_2$ through different channels. Interferences between electronic wave packets (evolving on attosecond timescales) are used to control the population of different electronic states of the excited neutral molecule, which can be switched on attosecond timescales. Then, by optimally selecting the excitation wavelengths and time delays, we can control the vibrational motion, total excitation, ionization yield and desired ionization and dissociation pathways. State-of-the-art quantum calculations, which have only recently become feasible, allowed us to interpret this very rich set of quantum dynamics, including both the nuclear motion *and* the coherently excited electronic state interferences. Thus, we succeed in both observing and rigorously modeling multi-scale coherent quantum control in the time domain for the first time. The observed richness and complexity of the dynamics, *even in this very simplest of molecules*, is both remarkable and daunting.

We note that our approach for control, using a combination of phase-locked VUV and IR fields, where the VUV field consists of an *attosecond pulse train* with a 10 fs pulse envelope, is ideal for coherently exciting and manipulating electronic and vibrational states on attosecond time scales, while simultaneously retaining excellent spectral resolution necessary for state-selective attochemistry. Exciting electron dynamics in a molecule using a single, broad-bandwidth, VUV attosecond pulse would simply excite many ionization/dissociation channels with little state selectivity, potentially masking the coherent electronic and nuclear quantum dynamics. For example, the bandwidth required to support an isolated 200 attosecond pulse around 15 eV is ≈ 5 eV. In contrast, the 10 fs VUV pulse train used here corresponds to a comb of VUV harmonics, each with a FWHM bandwidth of 183 meV, that can be tuned in the frequency domain (26) to coherently and selectively excite multiple electronic states.

## Experiment

Figure 1 illustrates our new concept for attosecond coherent control of molecular dynamics. A neutral deuterium molecule is electronically excited and ionized by combined phase-locked VUV harmonics (7ω – 13ω) and ultrafast IR (ω) pump fields at a center laser wavelength of 784 nm (see Supplementary Materials for more details). A second control IR pulse is time-delayed with respect to the combined VUV + IR fields. For simplicity, only the two most relevant electronically excited states of different parity in $D_2$ are shown in Fig. 1. As we will explain in more detail below, interferences between electron wave packets excited by the combined VUV harmonics (i.e. 7ω, 9ω, and 11ω) and the IR field (ω) are used to manipulate the ionization and excitation probabilities of different electronic states of the excited neutral molecule. To modulate the total electronic excitation in the neutral molecules on attosecond time scales, we use two-pathway quantum interference of electronic wave packets excited by 7ω+ω and 9ω–ω. Simultaneously, to modulate the total ionization yield on attosecond time scales, we use two-pathway quantum interference of electronic wave packets excited by 9ω+ω and 11ω–ω as the main mechanism. Finally, by tuning the VUV excitation wavelengths, we achieve other degrees of coherent control of coupled electron-nuclear wave packet dynamics, including vibration, ionization and tunneling. First, we show how the population of two electronic states can be switched (i.e. between EF an B) on attosecond time scales (Fig. 1 lower left). Second we show that by tuning the energy of the VUV harmonic comb, we can excite and control the population of different electronic and vibrational states, which in turn, dictates how the neutral molecule vibrates and ionizes. Third, by exciting $D_2$ using tunable two-color VUV and IR fields, and then probing the dynamics using a second IR pulse, we can control tunneling of nuclear wavepackets in the EF potential of a neutrally excited $D_2$.

In our experiment, the 7$^{th}$ and 9$^{th}$ harmonics coherently excite the molecule from its ground state, creating two nuclear wave packets in the same, odd-parity (i.e. B $^1\Sigma_u^+$) potential energy surface of $D_2$. When a small portion of the driving IR field (at an intensity of $3x10^{11}$ W/cm$^2$) co-propagates with the VUV harmonics, we can also simultaneously populate the optically forbidden, even-parity (i.e. EF $^1\Sigma_g^+$) potential energy surface through two-photon (i.e. $7\omega + \omega$, $9\omega - \omega$) absorption processes. A second, time-delayed, and stronger IR ($\omega$) pulse (at an intensity of $4x10^{12}$ W/cm$^2$) is then used in two different ways. First, on short attosecond time scales, the delayed IR field interferes with the IR field that co-propagates with the VUV harmonics, and serves as a knob to control the excitation and ionization processes on attosecond time scales in the Franck-Condon region, by means of electron wave packet interferometry. And second, on long fs timescales, the delayed IR field serves as a femtosecond knob to control the dissociation process by selectively ionizing the molecule at some optimal time after excitation. In this experiment, we used the COLTRIMS technique (*28*) to simultaneously collect the full 3D electron and ion momenta of all the reaction products which include $D_2^+$ and an electron, as well as the $D^+$ dissociative ions (see SM for more details). All the pulses (VUV attosecond pulse trains and two IR fields) were linearly polarized in the same direction. The VUV pulse duration was 5-10 fs, while the IR pulse durations were 30 fs. By changing the pressure in the gas-filled capillary, we can fine-tune the exact photon energies of the VUV harmonics while keeping the IR wavelength constant. This capability is key for uncovering the control mechanisms.

We first examine the attosecond control pathways at early times, before the onset of large-period vibrational wavepacket dynamics. Figure 1 (top left inset) plots the experimental photoelectron, $D_2^+$ and $D^+$ yields when the pump (VUV + IR) and the control (IR) pulses were overlapped in time, as a function of the time delay between them (see SM for additional data taken when the VUV harmonics were tuned to different photon energies). Very rapid, sub-optical-cycle, modulations in the photoelectron and $D_2^+$ yields result from the combination of optical interferences (two IR pulses) and quantum interferences of electronic wavepackets (i.e. $9\omega+\omega$ and $11\omega-\omega$ interfering pathways), as the phase of the control IR pulse changes relative to the pump VUV+IR pulse. Half-a-cycle periodicity suggests that the two-pathway quantum interferences play an important role in this particular case (*29*). The deep, full-cycle modulation of the dissociative $D^+$ yield is simply a result of the optical interferences between the pump and control IR pulses (both 30 fs long) that lead to bond-softening of the ground state of $D_2^+$ and ionization of the $D_2^*$ excited states approximately 10 fs after the pump pulse. Since the dissociation by bond-softening occurs after the VUV pulse is gone, this signal provides a reference point of the absolute phase in between the two IR pulses at each time delay, and allows us to precisely know the phase of the quantum interferences relative to the laser field.

Theoretical calculations (Fig. 1 lower left inset) show that the electronic population in the excited states also oscillates, with the same half-cycle periodicity. Moreover, theoretically we can remove the parity degeneracy in the total excitation yield and see that the populations in the gerade and ungerade B and EF states are out of phase. Thus, a two-pathway quantum interference of the electron wave packets driven by the lower two harmonics and the delayed IR field (i.e. $7\omega+\omega$ vs $9\omega-\omega$, lower inset Fig. 1) can be used as an ultrafast population switch between even (EF) and odd parity (B) potentials. *This demonstrates that the interference of electronic wavepackets can be used to switch and steer the electronically excited states on attosecond time scales, allowing simultaneous control of the electronic and vibrational* *excitation of an excited neutral $D_2^*$ molecule.* Thus, we demonstrate how combined attosecond VUV and IR femtosecond fields can be used as a novel tool to coherently control chemical reactions on the fastest timescales.

It is worth noting that while there are other possible interfering pathways responsible for the excitation and the ionization yield modulations, they are significantly less likely because they require more photons. For example, the interference of electronic wave packets excited by $7\omega+3\omega$ and $11\omega-\omega$ is possible. In this case, the total ionization probability is controlled by coupling the lower vibrational states of the B potential excited by the $7^{th}$ harmonic, with the continuum electron wave packets created by the $11^{th}$ harmonic. While this channel competes with the state excited by the $9^{th}$ harmonic, it requires absorption of four photons, and is thus much less probable compared with the pathway requiring absorption of two photons (one VUV and one IR).

**Theory**

Since we solve the full 3D time-dependent Schrödinger equation (TDSE) for $H_2$ exposed to a combined VUV and IR fields, we can theoretically examine the attosecond control mechanisms by fine-tuning the photon energy of the VUV pulse. The TDSE is numerically solved by expanding the time-dependent wave function in a large basis of Born-Oppenheimer (BO) molecular states, which are obtained by diagonalization of the electronic and nuclear Hamiltonians of $H_2$ (see Supplementary Information for more details). The method includes all electronic and nuclear degrees of freedom and, therefore, accounts for electron correlation and the coupling between the electronic and nuclear motions. Time evolution starting from the ground state induces transitions between the BO states through laser-molecule and potential couplings. To make these complex calculations tractable, in the simulation we used IR and VUV pulses of 7.75 fs total duration. Although these pulses were shorter than the ones used experimentally (30 fs and 5-10 fs respectively), a direct comparison between the theory and the experiment can still be made during the time-delay interval when the pump and probe pulses overlap. The theoretical results are shown in Figs. 2(*A*-*C*), and capture the main experimental observations – that the total yields modulate on attosecond timescales, and that the periodicity and the amplitudes of the oscillations strongly depend on the exact central energy of the VUV pulse. Here, we keep the IR laser wavelength constant while blue shifting the VUV central wavelength as though the higher harmonics were generated by 784 nm, 770 nm, and 760 nm laser wavelength (Figs. 2(*A*-*C*), respectively). Not surprisingly, the modulation of the total ionization yield strongly depends on the exact energy of the VUV harmonics since the phases and the amplitudes of the interfering electron wave packets strongly depend on the laser-modified electronic structure of the molecule (28). In the case of $H_2$, as the central energy of the VUV beam blue-shifts from harmonics of 784 nm to harmonics of 740 nm, the 9$^{th}$ harmonic does not excite the B state in the Franck-Condon region, but accesses higher electronic states (see SM for more detail). In this regime the Born-Oppenheimer approximation breaks down, since the evolution of the nuclear wavepackets in the B and EF potentials is on the same time scale as the duration of the attosecond VUV pulse train (Fig. 1 lower panel inset), and can influence the electron wave packet interference process as well. These results thus demonstrate that we can precisely control which electronic states are excited by the tunable VUV harmonics, and show how these states can be switched on attosecond time scales.

**Multiscale Quantum Control of Electronic and Nuclear Dynamics**

To show experimentally how tunable attosecond VUV pulse trains can be used to precisely control the excitation probabilities of different electronic and vibrational states, and how to steer ionization and dissociation in $D_2^*$, we delay the control IR field relative to the VUV+IR pump field on femtosecond time scales. Figure 3 illustrates several possible ionization and dissociation pathways and plots the energies of the three different attosecond VUV harmonic combs we used to coherently and simultaneously control electronic and vibrational excitation, as well as nuclear wave packet tunneling in the EF potential. The combined 7ω+ω, 9ω−ω and 9ω fields coherently populate the EF $^1\Sigma_g^+$ and B $^1\Sigma_u^+$ states, respectively. These nuclear wave packets oscillate with periods that depend on the exact energy of the VUV harmonics, and can couple the ground state to different vibronic bands of the B and EF potentials. The probe IR pulse can then ionize and dissociate $D_2^*$ through different channels as the electronically excited neutral molecule vibrates. Here we focus only on three channels that leave the most visible signature in the kinetic energy release (KER) spectrum of $D^+$, which we label as “two-step B”, “one-step B” and “two-step EF” for short (panels A, B, and C) in Fig. 3. In the two-step B case (Fig. 3*A*), the nuclear wave packet launched in the B state by the 9$^{th}$ harmonic is first probed in the inner classical turning point of the B potential energy curve by absorption of two IR photons, thus ionizing the Rydberg electron from the B potential and launching a second nuclear wave packet in the $1s\sigma_g$ state of $D_2^+$. When the nuclear wave packet reaches the outer classical turning point of the $1s\sigma_g$ potential energy curve (after ≈ 10 fs), absorption of another IR-photon, from the same probe IR pulse, leads to an efficient coupling of the bound $1s\sigma_g$ and the dissociative $2p\sigma_u$ states of the ion, thus leading to dissociation into D + $D^+$ by a total absorption of 2+1 i.e. 3 IR photons (Fig. 3*A*). Through this two-step channel, the molecule dissociates with a kinetic energy release of about 0.7-0.9 eV, typical for a bond-softening process.

In the one-step B case (Fig. 3*B*), the neutral $D_2^*$ molecule first stretches to internuclear distances well beyond 9 a.u., which results in an increase of the effective ionization potential. At the instant the nuclear wave packet reaches the outer classical turning point of the B potential, the ionization of the molecule requires absorption of 3 IR photons. Through this channel, the nuclear wave packet is *directly* coupled to the $2p\sigma_u$ dissociative continuum and the molecule dissociates with lower KER compared with the two-step B case. Due to the absorption of an extra IR photon at the outer turning point, a modulation of the ionization probability is expected as the excited molecule vibrates in the B potential. Finally, in the two-step EF case (Fig. 3*C*), the nuclear wave packet launched by two photon absorption ($7\omega+\omega$ and $9\omega-\omega$) is probed in the inner classical turning point of the EF potential energy curve by absorption of three IR photons, which generates a nuclear wave packet in the $1s\sigma_g$ potential, but with smaller probability compared with the two-step B case since the probability of observing this channel corresponds to a sequential absorption of 3+1 IR photons.

Figure 4 validates the rich opportunities for state selective, coherent excitation of multiple nuclear wave packet dynamics of neutrally excited $D_2$, as well as a possibility of controlling bond breaking and dissociation through the pathways illustrated in Figs. 3(*A*-*C*). Figure 4*A* plots the experimental $D^+$ KER as a function of the pump-probe delay, while Fig. 4*D* plots the corresponding 2D Fourier transform (FT). One can clearly see oscillation periods of ≈ 83 fs and 58 fs at a $D^+$ KER of 0.7-0.9 eV, corresponding to the vibrational wavepackets in the B and EF states, respectively, that are obtained through the two-step B and two-step EF mechanisms. The larger probability of the one-step B channel versus the two-step EF channel (see Fig. 4*D*), confirms the 3 versus 4 IR photon absorption mechanisms. Moreover, Fig. 4*A* shows that the two-step B channel (high KER) has a larger probability than the one-step B channel (low KER), due to a lower effective ionization potential of the excited molecule at shorter internuclear separation. Figure 4*A* also shows that the two-step B and the one-step B channels are dephased, since the nuclear wave packet is probed at different times in those channels. This allows for steering of a desired dissociative route by precisely timing the second IR laser pulse as the neutral molecule vibrates.

As seen in Fig. 1, to reach the outer turning point of the double-well EF potential and the vibrational period of about 58 fs, the nuclear wave packet needs to tunnel through the potential barrier, which decreases the probability of observing the EF nuclear wave packet dynamics. The calculations nicely confirm the observed periodicities of these multiple nuclear wave packet dynamics in $D_2^*$ (see Figs. 4*G* and 4*H*). In order to model the dynamics of $D_2$ on femtosecond timescales, we first solved the TDSE to calculate the excitation probabilities, and then propagated the corresponding nuclear wave packets in the given $D_2^*$ potentials (see the SM for more detail on how the calculations were done). For example, Fig. 4*G* clearly shows how the EF nuclear wave packet starts reflecting *and* tunneling through the inner barrier at about R=3 a.u. and 10 fs after the excitation. We also see that the nuclear wave packet tunnels with about 20% probability compared with the main wavepacket motion occurring in the inner well of the EF double potential well. Moreover, the other striking feature of this calculation starts at about 45 fs, when the nuclear wave packet in the inner well tunnels through the inner barrier for the second time and meets the outer-well nuclear wave packet on its way back from the outer classical turning point. The interference of the different nuclear wave packets is clearly visible from 45 fs onward. As the time evolves, multiple reflections/tunneling events from both sides of the inner barrier and the interferences of the nuclear wave packet in the EF potential are responsible for the fast decoherence of the EF nuclear wave packet. Movies, theoretical Fourier transforms, and further discussions illustrating dynamics in the $D_2^*$ are given in the Supplementary Materials.

Finally, by blue shifting the VUV harmonic wavelengths to effective driving wavelengths of 778 nm and 760 nm (while keeping the IR wavelength fixed at 784 nm), we can control the relative population of the different neutrally excited states of $D_2$. Figures 4*B* and 4*C* show the KER spectra as a function of pump-probe time delay, while Figs. 4*E* and 4*F* plot the corresponding 2D Fourier transforms. Two striking features are apparent. First, by slightly increasing the energies of the 7$^{th}$ and 9$^{th}$ harmonics, the vibration periods in the B and EF potentials simultaneously increase due to excitation to higher vibrational levels in the Franck-Condon region. And second, the relative strength of the one-step EF channel increases compared with the two-step B channel. The latter feature can be explained as follows. As seen in Figs. 1 and 3*C*, simultaneous absorption of $7\omega+\omega$ and $9\omega-\omega$ creates a nuclear wave packet that can tunnel through the inner potential barrier (located at R=3.5 a.u.) of the double-well EF potential. By slightly increasing the energy of the 7$^{th}$ harmonic, the tunneling process becomes more probable, thus increasing the nuclear wave packet density that vibrates with longer periods. This increases the relative visibility of the two-step EF channel with respect to the two-step B channel. For an effective driving wavelength of 760 nm, higher vibration levels of the EF potential are excited, allowing the nuclear wave packet to propagate freely above the inner-barrier, moving along the EF potential energy curve with a vibrational period that is now close to that observed in the B state. Again, this interpretation is confirmed by TDSE calculations given in the SM.

Finally, the two-step B channel also disappears when excited by 760 nm VUV harmonics, since the coupling of the ground and B states dramatically decreases in the Franck-Condon region. In this case, the 9$^{th}$ harmonic excites higher electronic states (i.e. B’ as shown in the supplementary material). Thus, we show how tunable attosecond VUV radiation can be used to precisely control excitation of different electronic and vibrational states.

## Conclusion

In conclusion, we present a powerful new approach for using tunable VUV and IR attosecond pulse trains to coherently excite and control an outcome of a simple chemical reaction in a $D_2$ molecule. The interference of electronic wavepackets excited by multicolor VUV and infrared fields can be used to control excitation and ionization on attosecond timescales, while maintaining good energy resolution and state selectivity. Selective bond breaking is achieved by controlling the excitation wavelength as well as the time delay between the pump and probe pulses. We also observe and control the nuclear wavepacket tunneling in the double-well EF potential. This work thus demonstrates broad new capabilities for doing attosecond chemistry.

## Acknowledgments

The authors gratefully acknowledge support from the Army Research Office and an NSF Physics Frontier Center. Work partially supported by the Advanced Grant of the European Research Council XCHEM 290853, the European grant MC-RG ATTOTREND, the European COST Actions CM0702 and CM1204, the European ITN CORINF, the MICINN project Nos. FIS2010-15127 and CSD 2007-00010 (Spain), and the ERA-Chemistry project PIM2010EEC-00751. XMT was supported by Grand-in-Aid for Scientific Researches (No. C24540421) from the Japan Society for the Promotion of Science and HA-PACS Project for advanced interdisciplinary computational sciences by exa-scale computing technology.

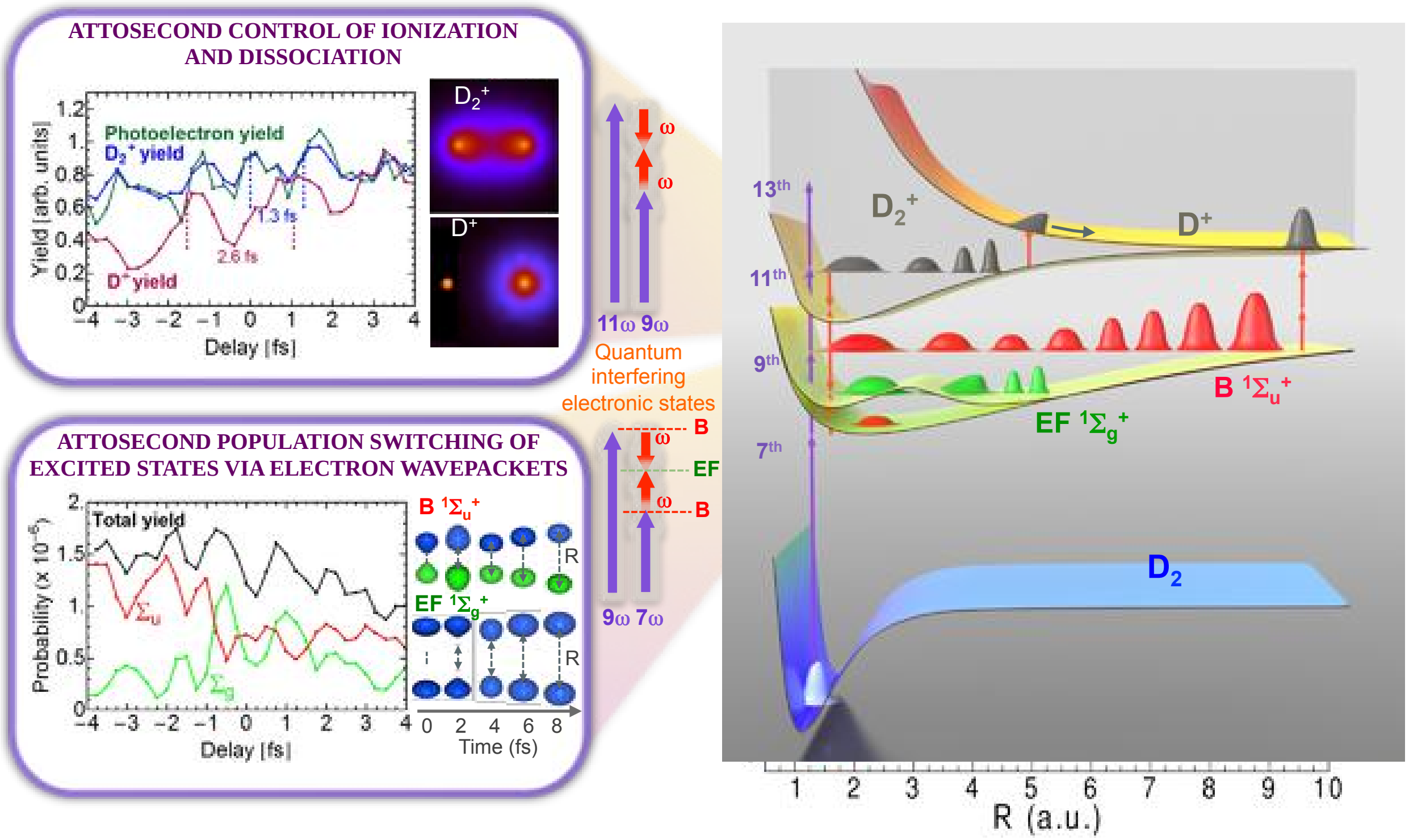

**Figure 1: Attosecond control of the excitation and ionization pathways of $D_2$ on short timescales. (right)** Simplified potential energy surfaces of the $D_2$ and $D_2^+$. The purple and red arrows represent the VUV harmonics and IR photons used to coherently control the populations of the excited neutral and ion states in the Franck-Condon region through two-pathway quantum interference of electronic wave packets in B, B* (single photon) and EF (two photon) states. (Lower left inset) Calculated excitation probabilities into states of $\Sigma_u$ and $\Sigma_g$ symmetries of neutral $H_2$ (dominated, respectively, by the B and EF states) as a function of time delay. The blue lobes, plotted on the right of the panel, are sketches of the $\Sigma_u$ and $\Sigma_g$ orbitals representing the excited electron dynamics. Theoretical predictions show that the electronically excited populations can be switched between the even B and odd parity EF potentials on attosecond timescales, which can in turn control how the molecule vibrates. (Upper left inset) The experimental photoelectron, $D_2^+$ and $D^+$ yields modulate on full and half-cycle attosecond timescales, as the delay between the pump VUV+IR and control IR pulses is scanned.

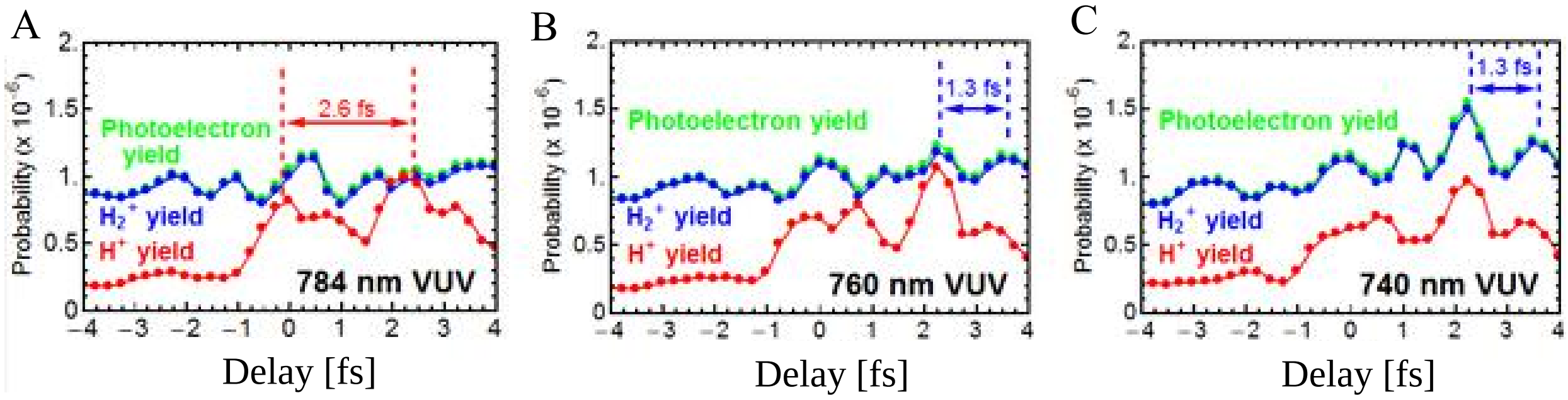

**Figure 2. Theory of attoseocond VUV coherent control of $H_2$ ionization/dissociation channels.** (*A-C*) As the central photon energy of the VUV harmonics is blue-shifted from an effective driving laser wavelength of 784 nm to 740 nm, the total ionization yield switches to half-cycle periodicity due to different electronic wavepacket interferences, thus demonstrating attosecond coherent control over the interfering electron wave packets and ionizing pathways in molecules.

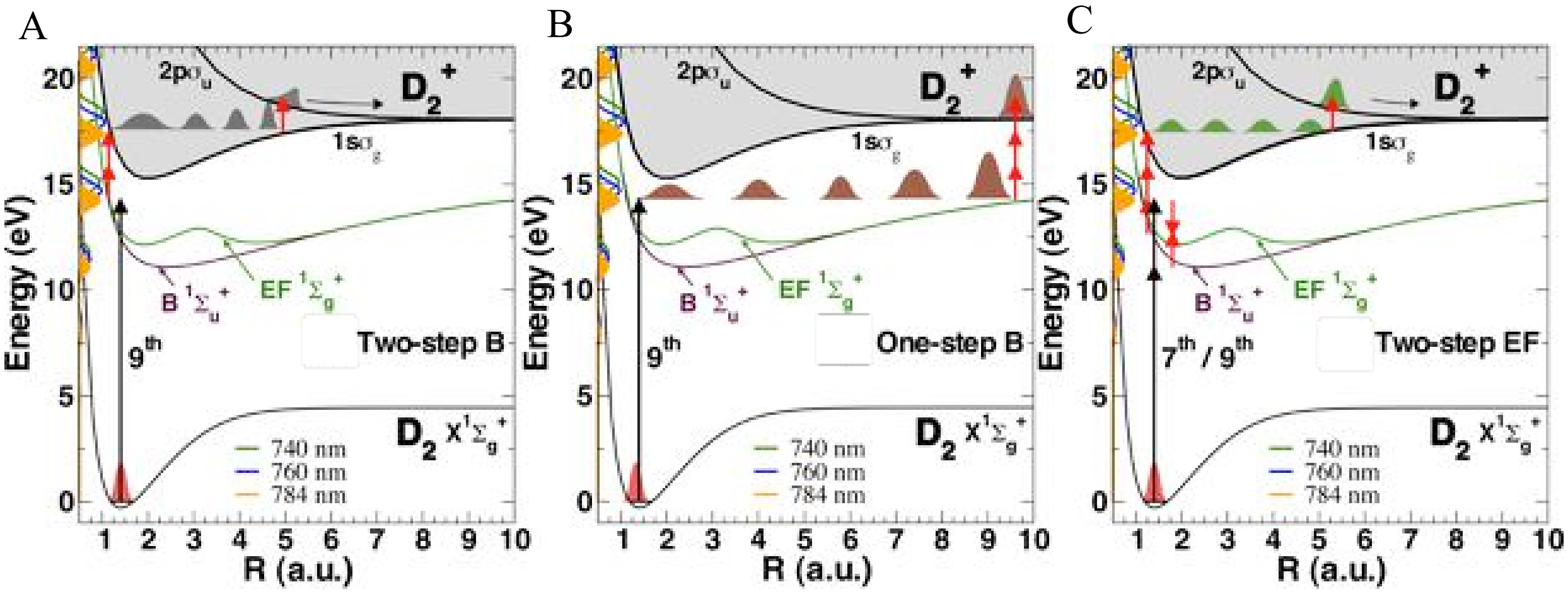

**Figure 3. Combined coherent control of electronic and nuclear wavepacket dynamics on longer timescales.** (*A*) Two-step B mechanism ($D^+$ high kinetic energy). (*B*) One-step B mechanism ($D^+$ low kinetic energy). (*C*) Two-step EF mechanism ($D^+$ high kinetic energy). We also show the tuning range of the VUV harmonics on the y axis (i.e. effective driving laser wavelengths of 784 nm, 760 nm, and 740 nm).

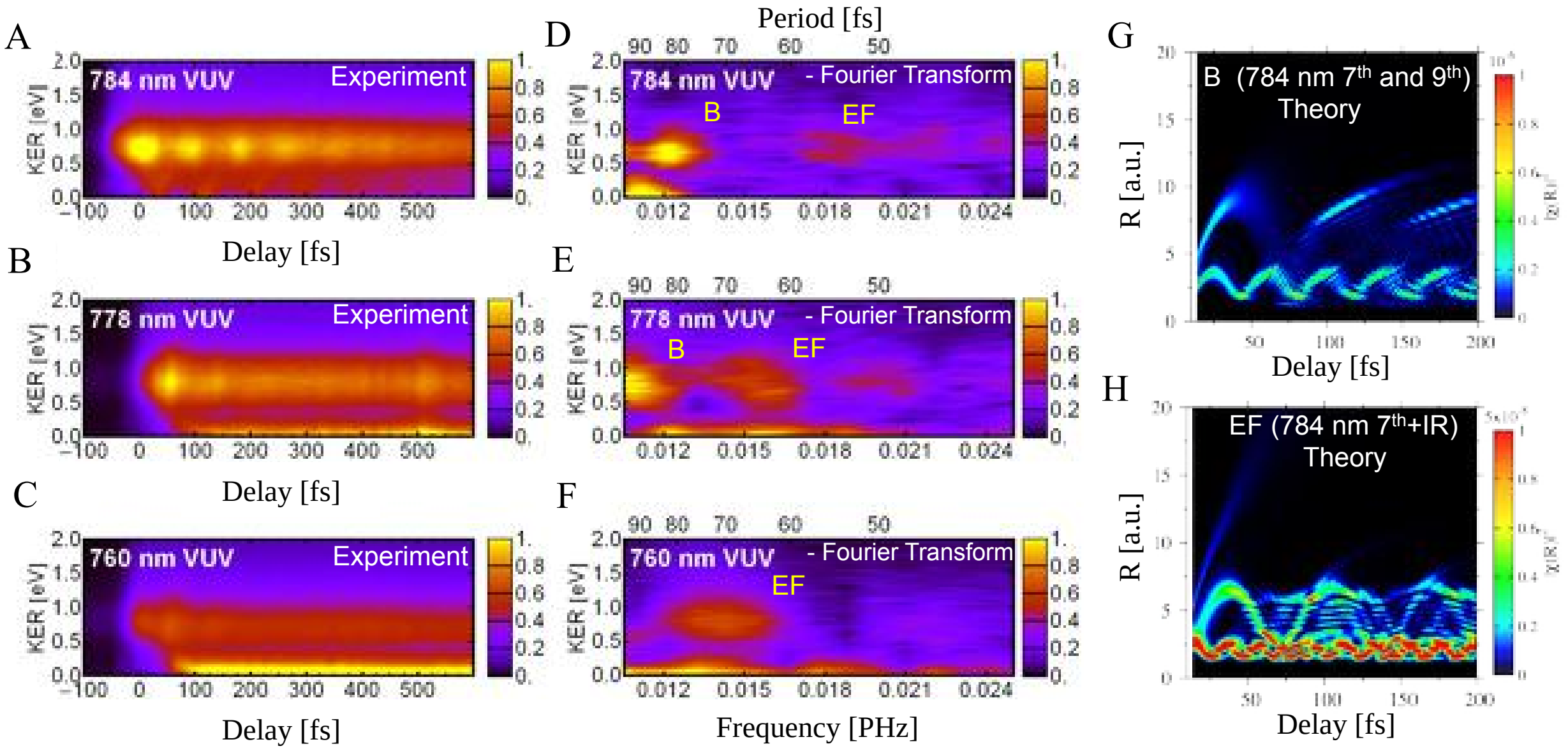

**Figure 4. Controlling the dissociation channels by tuning the photon energy and through nuclear wave packet dynamics.** (*A*-*C*) $D^+$ kinetic energy release resulting from dissociation using three different VUV frequency combs, that can be tuned to excite different $D_2^*$ electronic and vibrational states. (*D*-*F*) Corresponding 2D Fourier transforms of the $D^+$ kinetic energy releases. (*G*) Calculated free evolution of the nuclear wavepackets generated by the combined VUV + IR pulses in the B potential energy surface. (*H*) The same for the EF potential energy surface.

# *Supporting Information*

# Attosecond VUV Coherent Control of Molecular Dynamics

P. Ranitovic[1,4], C. W. Hogle[1], P. Rivière[3], A Palacios[3], X. M. Tong[2], N. Toshima[2], A. González-Castrillo[3], L. Martin[1], F. Martín[3,5], M. M. Murnane[1], and H. C. Kapteyn[1]

[1]*JILA and Department of Physics, University of Colorado and NIST, Boulder, CO 80309, USA*

[2]*Division of Materials Science, Faculty of Pure and Applied Science, University of Tsukuba, Ibaraki 305-8573, Japan*

[3]*Departamento de Química, Universidad Autónoma de Madrid, 28049 Madrid, Spain*

[4]*Lawrence Berkeley National Lab, 1 Cyclotron Road, Berkeley CA, 94720*

[5]*Instituto Madrileño de Estudios Avanzados en Nanociencia, IMDEA-Nano, Cantoblanco, 28049 Madrid, Spain*

In this supporting text, we first provide additional details about the experimental setup. Next, we provide additional data to explain in detail how the coupled electronic wavepacket, IR laser interferences, and nuclear wavepacket dynamics allow us control excitation and dissociation channels in $D_2$. Finally, we explain our fully quantum calculations that model the short timescale electronic wavepacket interference dynamics in $H_2$. We also provide additional theory on the longer timescale dynamics that beautifully show that we can exquisitely control tunneling and oscillation in a double-well excited state neutral molecule potential, by tuning the HHG photon energies and electronic wavepacket interferences.

## Experiment

Our experimental setup consists of a high-power (30 W), high repetition rate (10 kHz) 30 fs Ti:sapphire laser system (784 nm), a gas-filled waveguide for generating harmonics, and a COLTRIMS apparatus which allows for simultaneous detection of ion and electron 3D momenta. Using part of the laser output, high harmonics (HHG) were generated in Xe gas and then refocused into a separate $D_2$ gas target using a pair of multilayer VUV mirrors that reflect photon energies up to 22 eV. (To achieve the maximum efficiency, the supersonic COLTRIMS $D_2$ jet was replaced with an effusive, thick target.) The exact energies of the harmonics can be continuously tuned by the phase-matching conditions via pressure tuning of Xe, as explained below. A small portion of the IR pulse that drives the HHG process co-propagated with the VUV pulse. These two pulses (VUV/IR) with similar intensities were locked in phase to form one multicolor pump beam. The probe IR beam was delayed relative to the pump beam using a picomotor-driven delay stage. This VUV/IR interferometer was interferometrically stable to enable the interference of the electronic wavepackets. The VUV and IR pulse durations were 10 fs and 30 fs, respectively, while the pump and probe IR pulse intensities were $3x10^{11}$W/cm$^2$ and $4x10^{12}$ W/cm$^2$, respectively. All the pulses were linearly polarized.

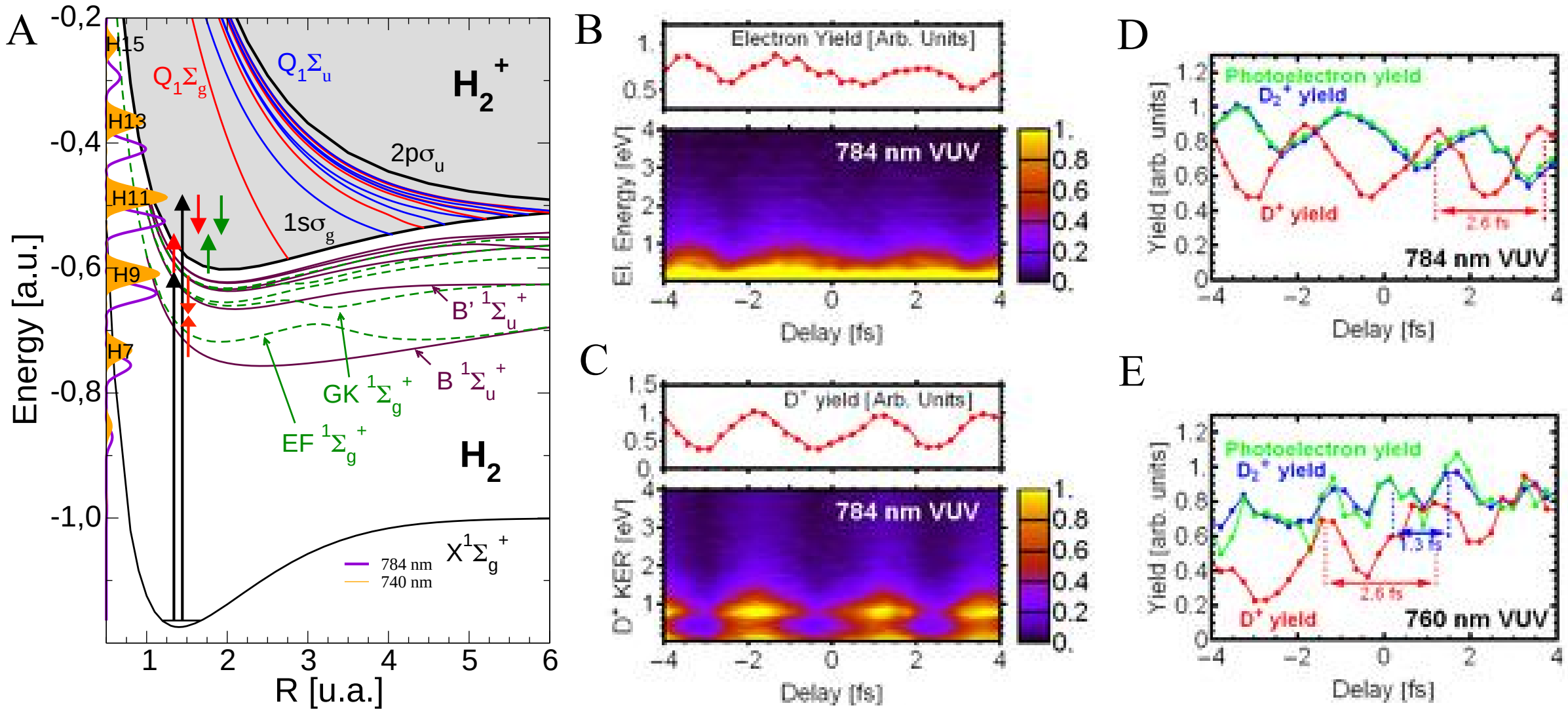

**Figure S1:** ***A)*** Detailed potential energy surfaces (PES) of the $H_2$ neutral and the $H_2^+$ ground and excited states (identical to those of $D_2$ and $D_2^+$). The red and green arrows represent driving and probing IR pulses, respectively. ***B)*** The photoelectron yield modulation results from the laser-induced interference of the electron wave packets created by the attosecond VUV frequency comb. The IR laser wavelength was 784 nm, while the VUV photons were 7ω, 9 ω, 11 ω, etc.. ***C)*** The $D^+$ dissociative yield modulation results from the intensity modulation of the interfering driving and probing IR pulses. ***D)*** A comparison of the $D^+/e^-$ yields reveals that the ionization and dissociative dynamics are out of phase. ***E)*** By increasing the energy of the VUV photons, we change the electron-yield modulation period from full to half-optical-cycle periodicity.

Figure S1 is an extended version of Fig. 1 from the main manuscript and shows in more detail ground and excited states of a deuterated hydrogen molecule ($D_2$ is the same as $H_2$). We also show the photoelectron, $D^+$ and $D_2^+$ yields, as well as kinetic energies of the electrons and the $D^+$ ions, for two different central wavelengths of the VUV frequency comb. The wavelength (photon energy) of the VUV harmonics was fine-tuned while keeping the driving and probing IR wavelengths unchanged by adjusting the Xe pressure (and thus the HHG phase matching) in the gas-filled waveguide[1-3], as shown in Fig. S2. As the Xe pressure in the HHG waveguide was increased from 5 to 20 torr, the photoelectron energies that result from ionizing Ar by the VUV HHG comb also increased because it is possible to phase match at slightly higher photon energies. This ability to fine-tune VUV photon energies allows us to distinguish between various excitation and control mechanisms.

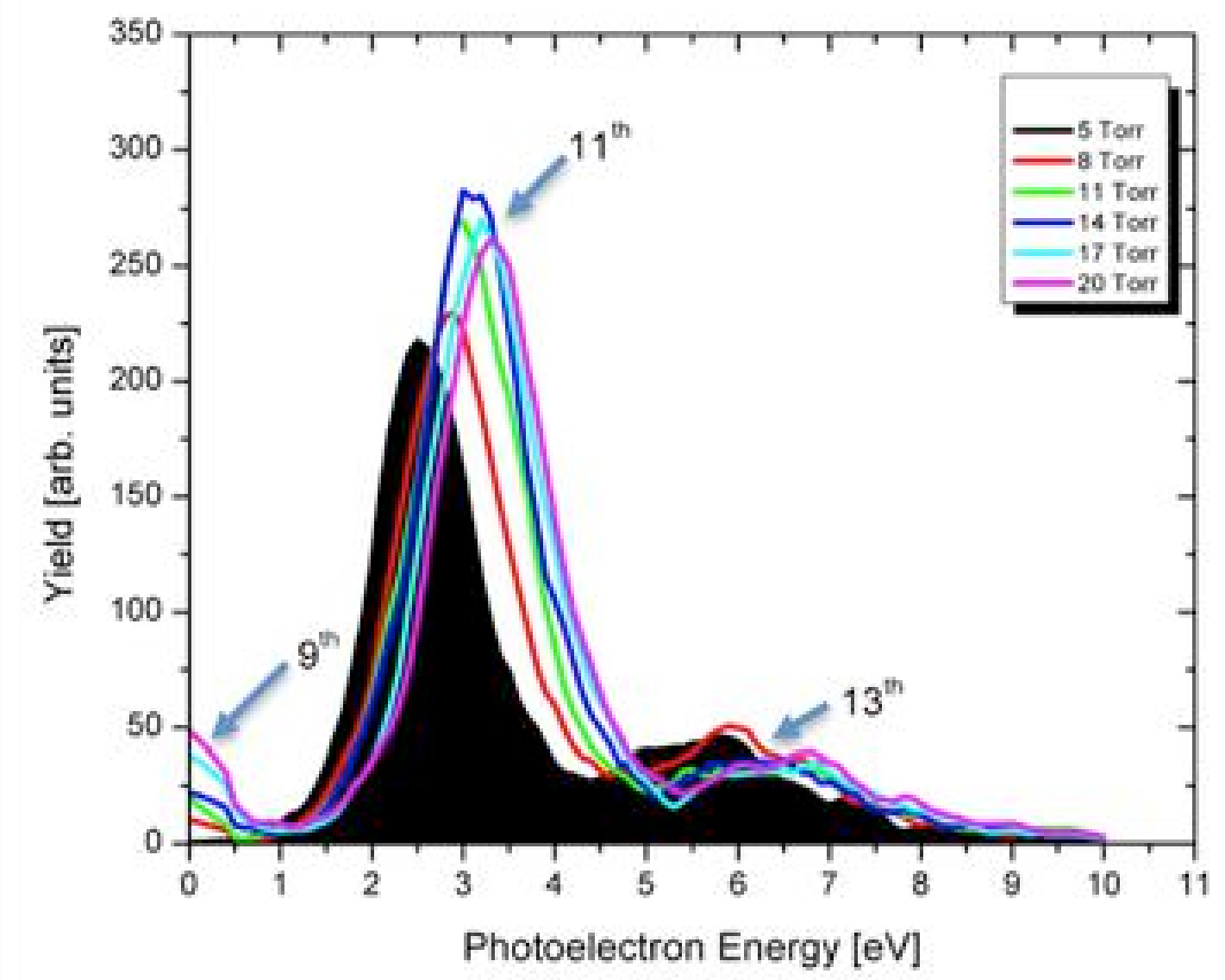

**Figure S2: Fine tuning of the VUV HHG energy by phase matching.** Photoelectron energies from Ar ionized by a VUV HHG comb show how increasing the Xe gas pressure can blue shift the HHG peaks while keeping the IR wavelength fixed.

Figure 1*E* shows how this phase-matching approach allows for fine control of the periodicity of the $e^-/D_2^+$ yields that changes from full to half-optical-cycle duration of the IR laser field, as the harmonic energies were blue-shifted. We also see from Fig. 1 that the periodicity of the $D^+$ yield does not depend on the VUV photon energy. This is expected since the excited molecule was ionized in the combined field of the two IR pump and probe pulses, which determines the periodicity observed in the ion yield. Thus, the phase of the $D^+$ oscillation provides a reference for the instantaneous value of the total electric field that drives the dissociation of the bound $1s\sigma_g$ nuclear wavepacket (high kinetic energy release (KER) channel), as well as the dissociation through the low KER channel. While the high KER channel originates from a bond-softening process (which is routinely observed in many strong-field processes), the low KER channel originates from direct dissociation of $D_2$ as the highly excited Rydberg states were ionized at larger internuclear separations by the tail of the two IR pulses.

## Theory

Due to extremely lengthy and time-consuming theoretical calculations, we modeled pulses of duration 7.75 fs, for time delays no longer than 10 fs. In this calculation, we used an electric field consisting of a combined VUV and IR pump field followed by an IR probe field. Both IR fields had a frequency $\omega$ = 1.6 eV (775 nm) and were 3 cycles in duration. The pump IR pulse was phase-locked to the VUV pulse, with an intensity of $0.3 \times 10^{12}$ W/cm$^2$. The probe IR pulse intensity was $5 \times 10^{12}$ W/cm$^2$, corresponding to the estimated experimental IR intensities.

The VUV field consists of four pulses, with an intensity of $10^9$ W/cm$^2$ for the two central pulses and $0.35\times10^9$ W/cm$^2$ for the other pulses, which corresponds to an envelope with FWHM=3 fs and I=$1.14\times10^9$ W/cm$^2$. To simulate the blue-shift effect of the experimental VUV pulses, we used the VUV combs with the central frequency corresponding to 9$\omega$ for three different IR wavelengths: 784, 760 and 740 nm (or 1.58, 1.63 and 1.68 eV photon energies). The delay between the (VUV+IR) pump and the IR probe pulses was scanned from -4 fs to 4 fs (the pump precedes the probe for positive time delays).

We solved the time-dependent Schrödinger Equation (TDSE) for $H_2$ using a close-coupling method that includes the bound states, the $^2\Sigma_g^+(1s\sigma_g)$ and the $^2\Sigma_u^+(2p\sigma_u)$ ionization continua, and the doubly excited states embedded in them. The method is similar to that described in[4-6] and employed in[7,8]. The calculations consider all electronic and vibrational degrees of freedom, as well as the effects of electron correlation and interferences in between different ionization and dissociation channels. The field-molecule interaction is described within the dipole approximation, which is valid for the wavelengths employed here. In brief, we numerically solve the TDSE by expanding the time-dependent wave function in a basis of the vibronic eigenstates of the molecule ($H_2$ or $D_2$). The vibronic states are obtained within the Born-Oppenheimer approximation, i.e., they are written as products of electronic and vibrational wave functions. The molecular Hamiltonian and the laser field couple these states when the expansion is introduced in the TDSE. The bound and continuum electronic states are described by using a Feshbach-like method (see details in[4-6] and references there in). In this method, the electronic states results form configuration interaction (CI) calculations restricted to different subspaces. Each configuration in the CI expansion is written as an antisymmetrized product of $H_2^+$ orbitals described as monocentric expansions in terms of B-splines and spherical harmonics. In the present problem, the vibrational (dissociative) wave functions are represented in a basis of 300 B-splines inside a box of 14 a.u., and the radial part of the molecular orbitals in a basis of 180 B-splines inside a box of 60 a.u. and spherical harmonics up to l=16. We are focused on the detection of the fragments from the molecules aligned parallel to the laser polarization axis, so our present calculations only include states of Σ symmetry.

To model the NWP motion generated by the combined VUV+phase-locked IR pulse on a given electronic state (Figs. 4*A*, 4*B* and S4 and S5), we solved the nuclear TDSE by using the potential energy associated to that electronic state and the initial boundary condition:

$$\psi(R, t = t_0) = \sum_k c_k \chi_k(R) \exp(-i\varepsilon_k t_0)$$

where $c_k$ is the amplitude resulting from TDSE calculations as those described above for $H_2$, but in which only the combined VUV+phase-locked IR pulse is used (i.e. no IR probe), $\chi_k$ is the *k*-th vibrational wave function in the chosen electronic state, $\varepsilon_k$ is the corresponding vibronic energy, and $t_0$ is the time just at the end of the combined VUV+phase-locked IR pulse. To account for the fact that the $D_2$ reduced mass is twice as large as that of $H_2$, the time in the above equation has been rescaled by a factor $\sqrt{2}$.

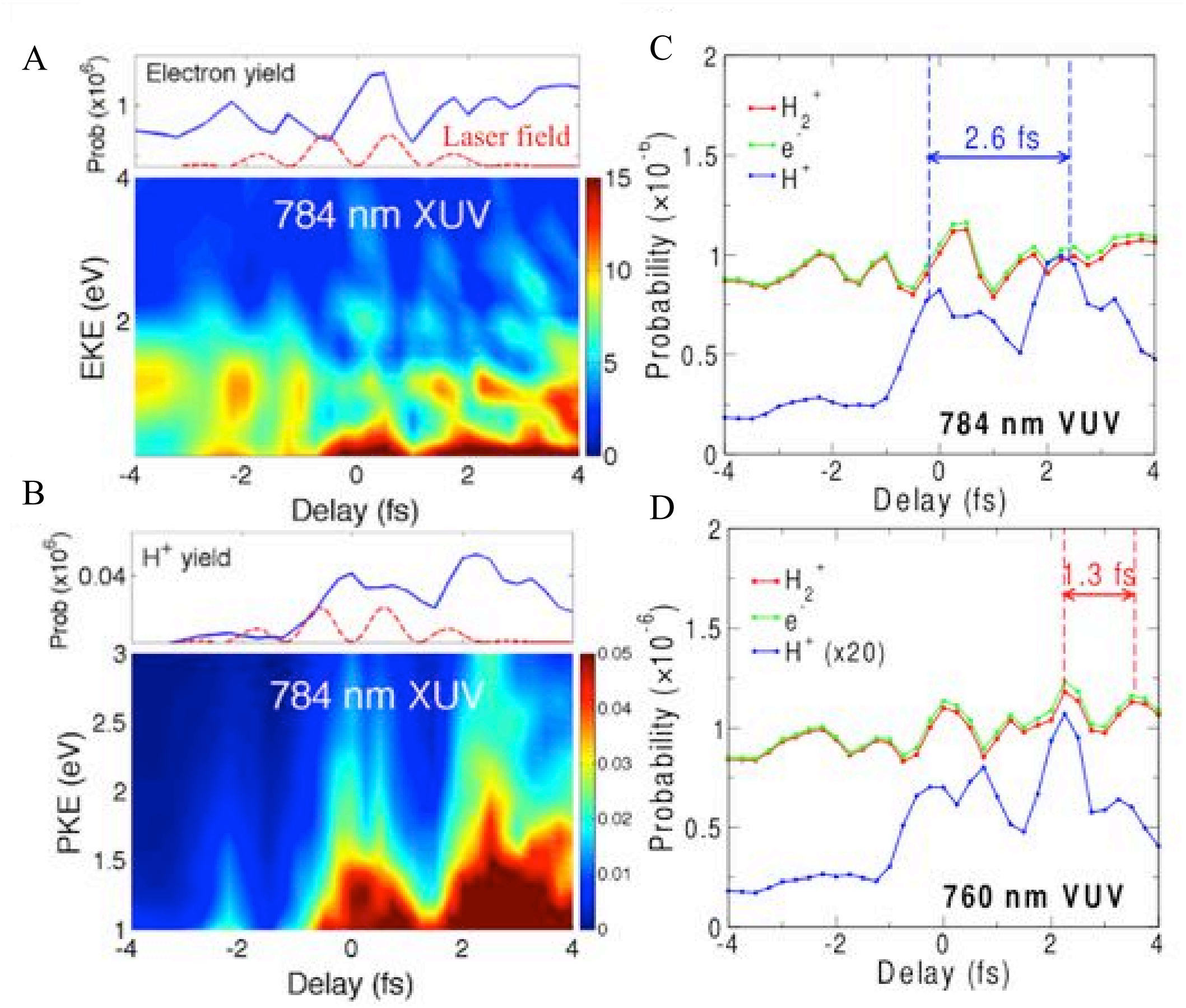

**Figure S3:** Calculated electron ***A)*** and proton ***B)*** kinetic energies (PKE) obtained by solving the TDSE for a hydrogen molecule in the presence of 5 fs VUV and IR pulses. ***C) and D)*** The calculated $H_2^+/e^-$ yields versus $H^+$ yields show the same trend as seen in the experimental data. Namely, as the central energy of the VUV frequency comb is blue-shifted from 784 nm to 760 nm, the total ionization yield clearly switches from full to half-a-cycle periodicity.

Figure S3 represents a more detailed version of Fig. 2 from the main text, showing the theoretical yields as well as the KE of the $D^+$ ions and the electrons. The photoelectron energies shown in Fig. S3 *A)* confirm that the two interfering electronic wave-packets (9ω+ω vs 11ω-ω) are indeed responsible for the total yield oscillation with the main periodical signal being just above the ionization threshold (10ω). Also, the proton kinetic energy shown in Fig. S3 *B)*, indicates that due to the shorter IR/VUV pulses used in our calculations, the dissociation happens at shorter internuclear separations (higher dissociation kinetic energy), where the pulses are still temporally overlapped, and where quantum interferences play an important role.

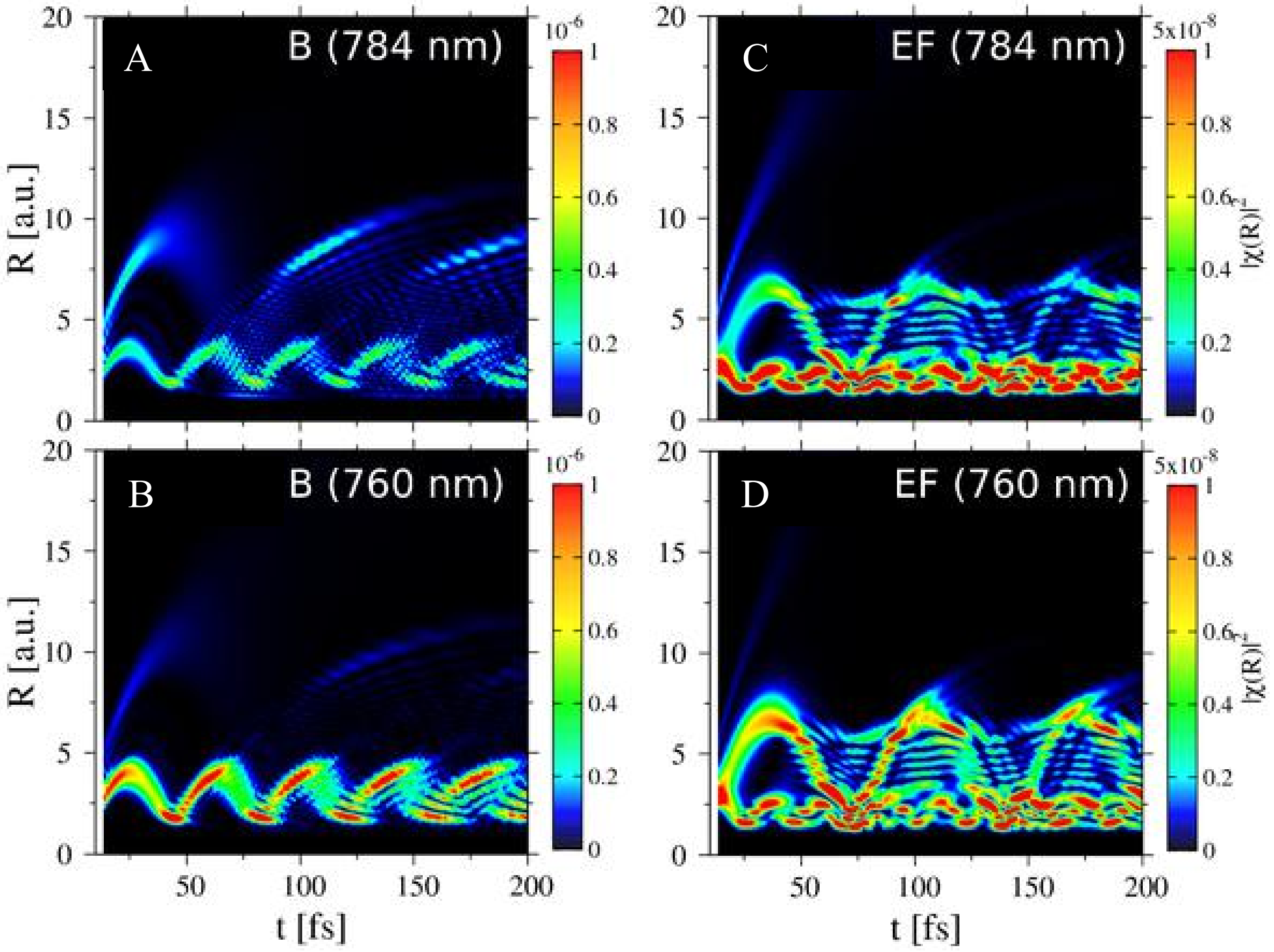

**Figure S4**: Time evolution of the nuclear wavepackets created in the B $^1\Sigma_u^+$ **(*A* and *B*)** and EF $^1\Sigma_g^+$ potential energy surfaces **(*C* and *D*)** by the combined VUV and IR fields, for 784 nm (top) and 760 nm (bottom) IR wavelengths.

In Figs. S4 and S5, we show the nuclear wavepacket dynamics calculations for the VUV central energies shown in Fig. 4, but for several other excited states populated by the VUV harmonics. For the B state at 784 nm (Fig. S4 *A*), two periodicities appear: ≈ 30 fs (small internuclear separation R), corresponding to population of low vibrational states of the B state by 7ω, and ≈ 80 fs (large R), which is probed by the one-step B mechanism. When the photon energy increases (i.e. 760 nm, Fig. S4 *B*), the vibration in the B potential at large R has a larger period due to excitation to higher vibrational levels in the Franck-Condon region, and its probability is smaller due to dissociation.

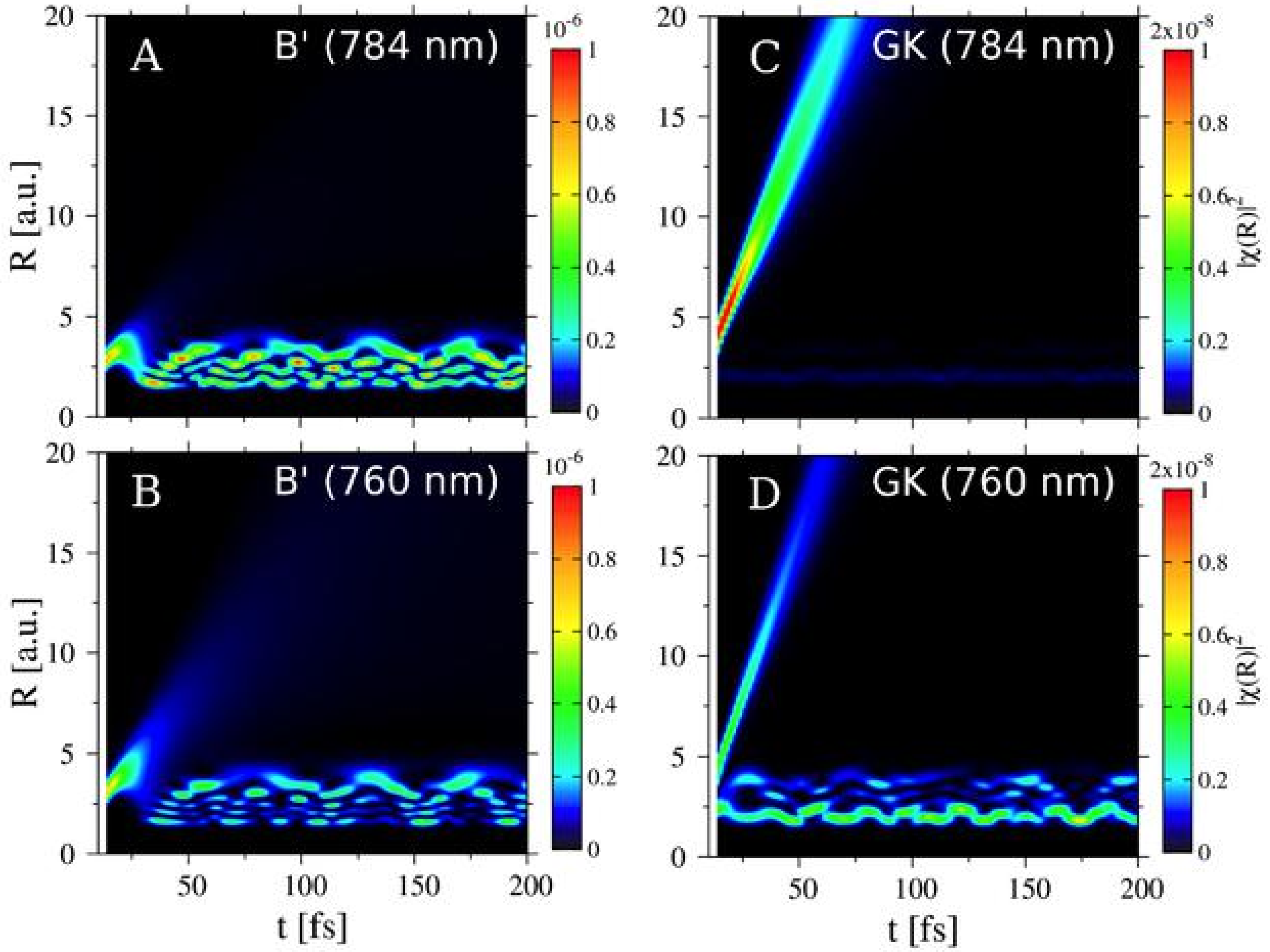

**Figure S5**: Same as in Fig. S4 but for the B' **(*A* and *B*)** and GK **(*C* and *D*)** states. The lowest vibrational states of the B' state are populated by the 9th harmonic (*A*). For higher photon energies (760 nm) the VUV absorption leads to dissociation (*B*). Excitation to the GK state leads primarily to dissociation (*C*), while for higher frequency the low vibrational states are also populated (*D*). Note that the populations are 50 times smaller in *C* and *D* than in *A* and *B*.

To observe these periodicities, we performed a Fourier transform of the time evolution of the nuclear wave packets shown in Fig. S4. The results are shown in Fig. S6. The predicted periodicities around 30 and 80 fs are present for the B state. A zoom over long internuclear distances of the Fourier transforms of the B state (shown in Figs. S6 *A* and *B*) is shown in Fig. S7. The vibration in the EF potential can also be seen at 784 nm (Figs. S4 *C*) and S6 *C*), but it is even more clearly visible for 760 nm wavelengths (Fig. S4 *D* and S6 *D*). This is because, at 784 nm, the nuclear wavepacket is mainly created in the inner and outer wells of the EF potential energy curve, with very little probability of tunneling through the potential barrier located at R ~ 3.5 a.u. Thus the corresponding vibrational periods are ≈ 20-30 fs at small R. In contrast, at 760 nm (Fig. S4 *D*) the nuclear wavepacket tunnels efficiently through the barrier and moves along the EF potential energy curve, with a vibrational period that is close to that observed in the B state. Note that the populations are 20 times smaller in Figs. S4 *C* and *D*, than in *A* and *B*.

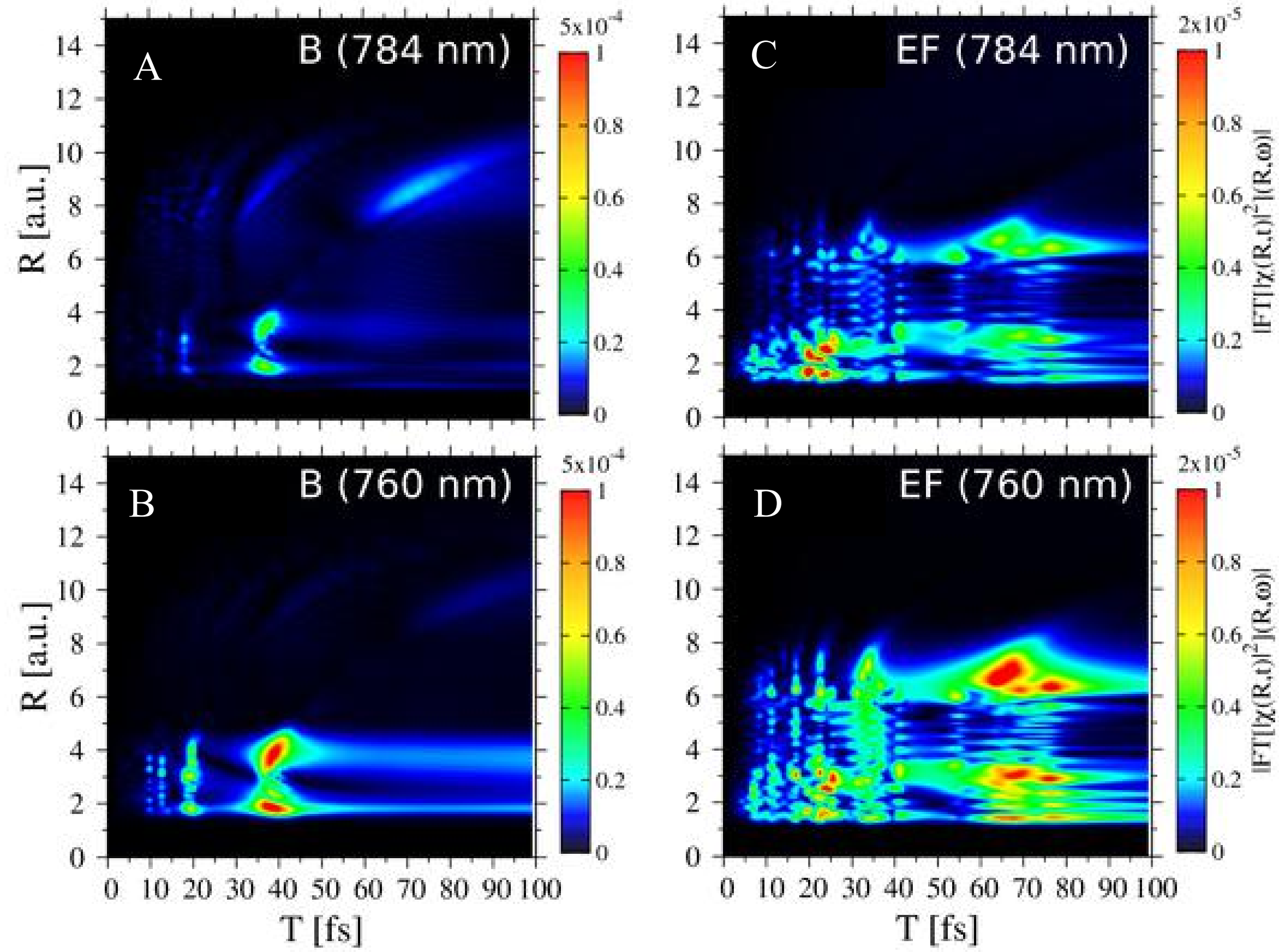

**Figure S6:** 2D Fourier transforms of the nuclear wave packets shown in Fig. S4.

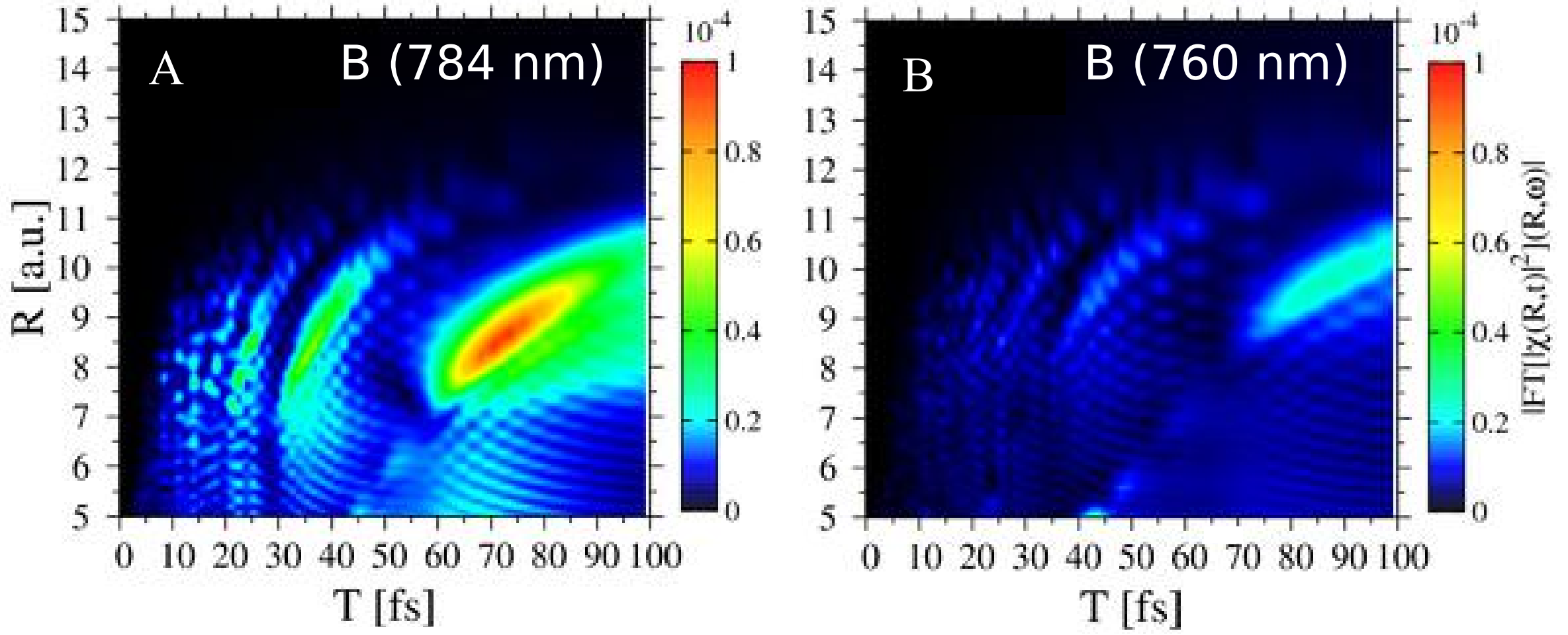

**Figure S7:** Zoom in for large internuclear distances of the 2D Fourier transforms of the nuclear wave packets shown in Fig. S6 *A* and *B*.

Finally, in Fig. S8 we show the extended Fourier transforms, corresponding to the data shown in Fig. 4 of the main manuscript. All structures associated with periodicities larger than 40 fs are explained in the caption of Fig. 4 and in the main text. Figure S7 also shows the appearance of peaks associated with periodicities in the range 20-40 fs, which correspond to nuclear wavepackets generated by the 7$^{th}$ harmonic in the B state, by the $7\omega + \omega$ and $9\omega - \omega$ in the inner (and likely outer) wells of the EF state, and by the $11\omega$ and $13\omega$ on the $1s\sigma_g$ state of $D_2^+$.

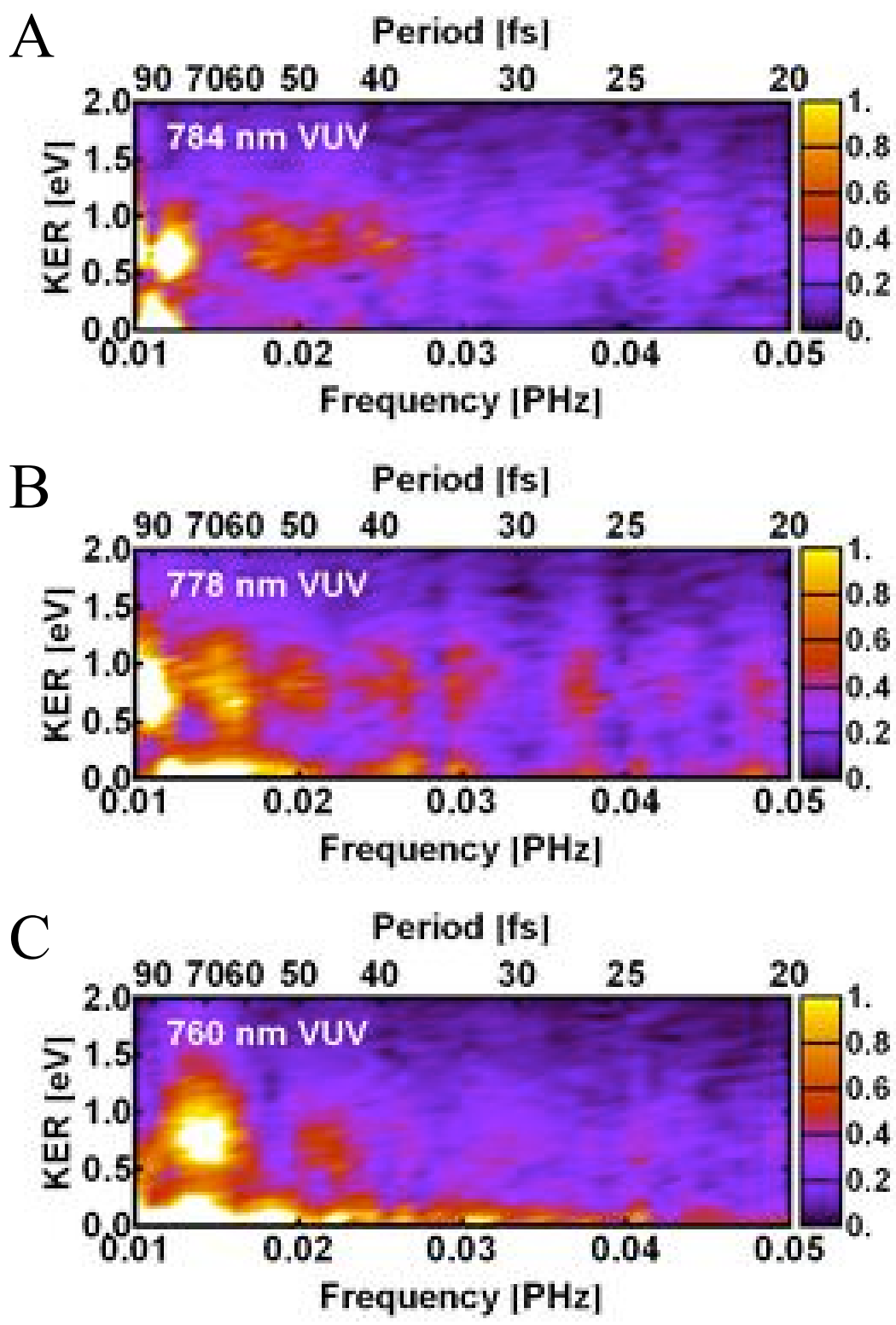

**Figure S8:** Extended 2D Fourier transforms of the proton kinetic energy releases shown in Fig. 4 *F-H* of the main text, as a function of the central energy of the VUV pulse.